\documentclass[10pt]{article}
\usepackage[textheight=9in,textwidth=6.5in]{geometry}

\usepackage{subfigure,lscape,color,graphicx,multicol}
\usepackage{url}

\def\Tabperfsc#1{\def\tabperfsc{#1}}
\Tabperfsc{1 of the Supplemental text}
\def\TabperfcritB#1{\def\tabperfcritB{#1}}
\TabperfcritB{2 of the Supplemental text}
\def\Tabgofnssh#1{\def\tabgofnssh{#1}}
\Tabgofnssh{3 of the Supplemental text}
\def\Tabtransfersh#1{\def\tabtransfersh{#1}}
\Tabtransfersh{4 of the Supplemental text}
\def\TabgofnsshANDtabtransfersh#1{\def\tabgofnsshANDtabtransfersh{#1}}
\TabgofnsshANDtabtransfersh{3 and 4 of the Supplemental text}

\begin{document}

\title{Comparing Protein Interaction Networks via a 
         Graph {\em Match-and-Split} Algorithm\thanks{Supplemental text is available at {\tt http://www.cs.berkeley.edu/{\textasciitilde}nmani/mas-supplement.pdf} .}}
\author{Manikandan Narayanan$^1$ and Richard M. Karp$^{1,2}$\\
{\small $^1$Computer Science Division, University of California, Berkeley, CA 94720}\\ 
{\small $^2$International Computer Science Institute, Berkeley, CA 94704}\\
{\small Email: \{nmani, karp\}@cs.berkeley.edu}}
\date{Nov 24, 2006}

\maketitle

\section*{Abstract}
  We present a method that compares the protein interaction networks of two
species to detect functionally similar (conserved) protein modules between
them. The method is based on an algorithm we developed to identify matching
subgraphs between two graphs. Unlike previous network comparison methods, our
algorithm has provable guarantees on correctness and efficiency.  Our algorithm
framework also admits quite general connectivity and local matching criteria
that define when two subgraphs match and constitute a conserved module.  

  We apply our method to pairwise comparisons of the yeast protein network with
the human, fruit fly and nematode worm protein networks, using a lenient
criterion based on connectedness and matching edges, coupled with a betweenness
clustering heuristic. We evaluate the detected conserved modules against
reference yeast protein complexes using sensitivity and specificity measures.
  In these evaluations, our method performs competitively with and sometimes
better than two previous network comparison methods. Further under some
conditions (proper homolog and species selection), our method performs better
than a popular single-species clustering method. 
  Beyond these evaluations, we discuss the biology of a couple of conserved
modules detected by our method. We demonstrate the utility of network
comparison for transferring annotations from yeast proteins to human ones, and
validate the predicted annotations.

\section{Introduction} \label{sec:intro}

\begin{figure}[bht]
\begin{minipage}{.45\textwidth}
  \includegraphics[width=2in]{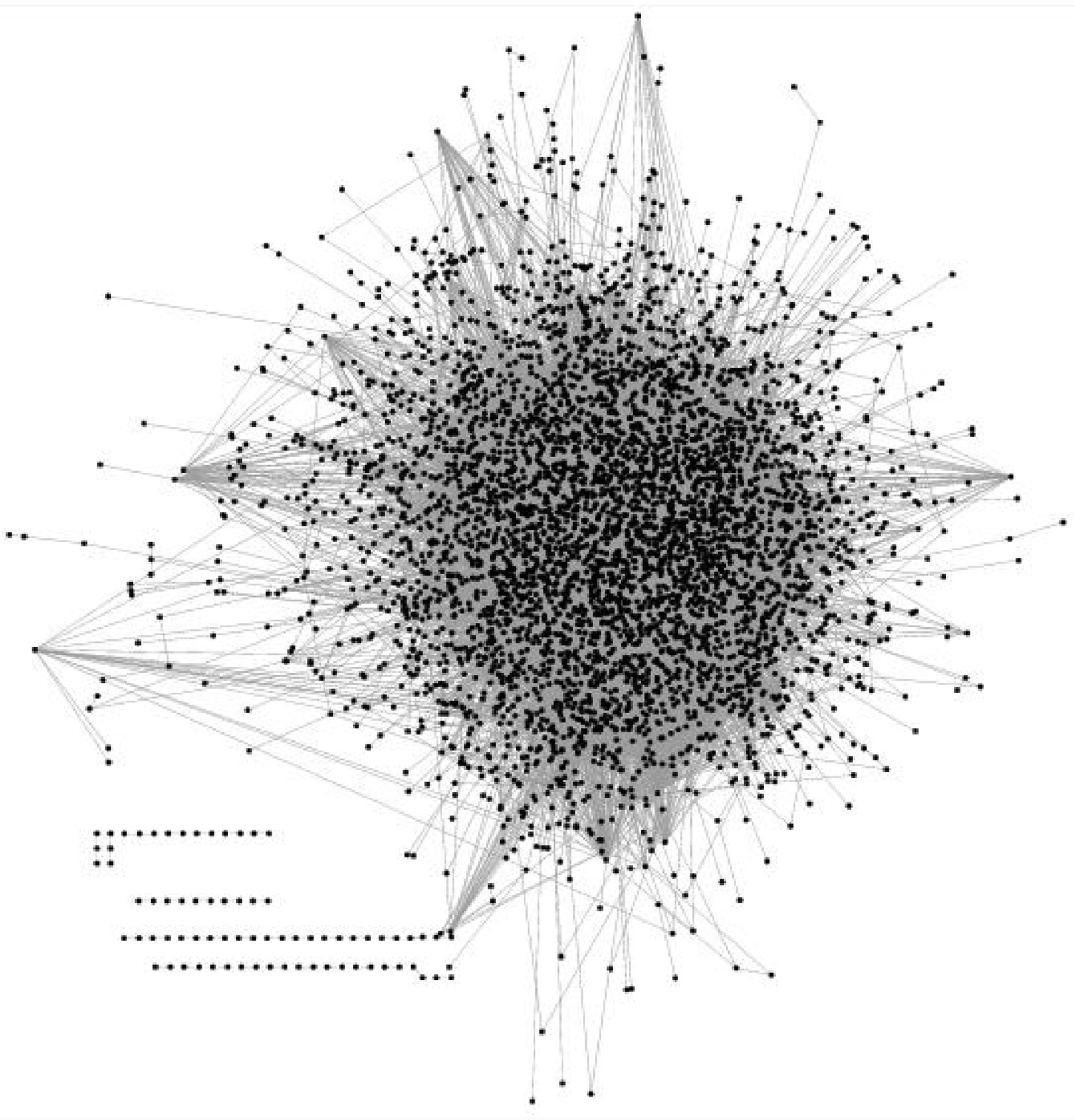}
\end{minipage}
\begin{minipage}{.45\textwidth}
  \includegraphics[width=2.5in]{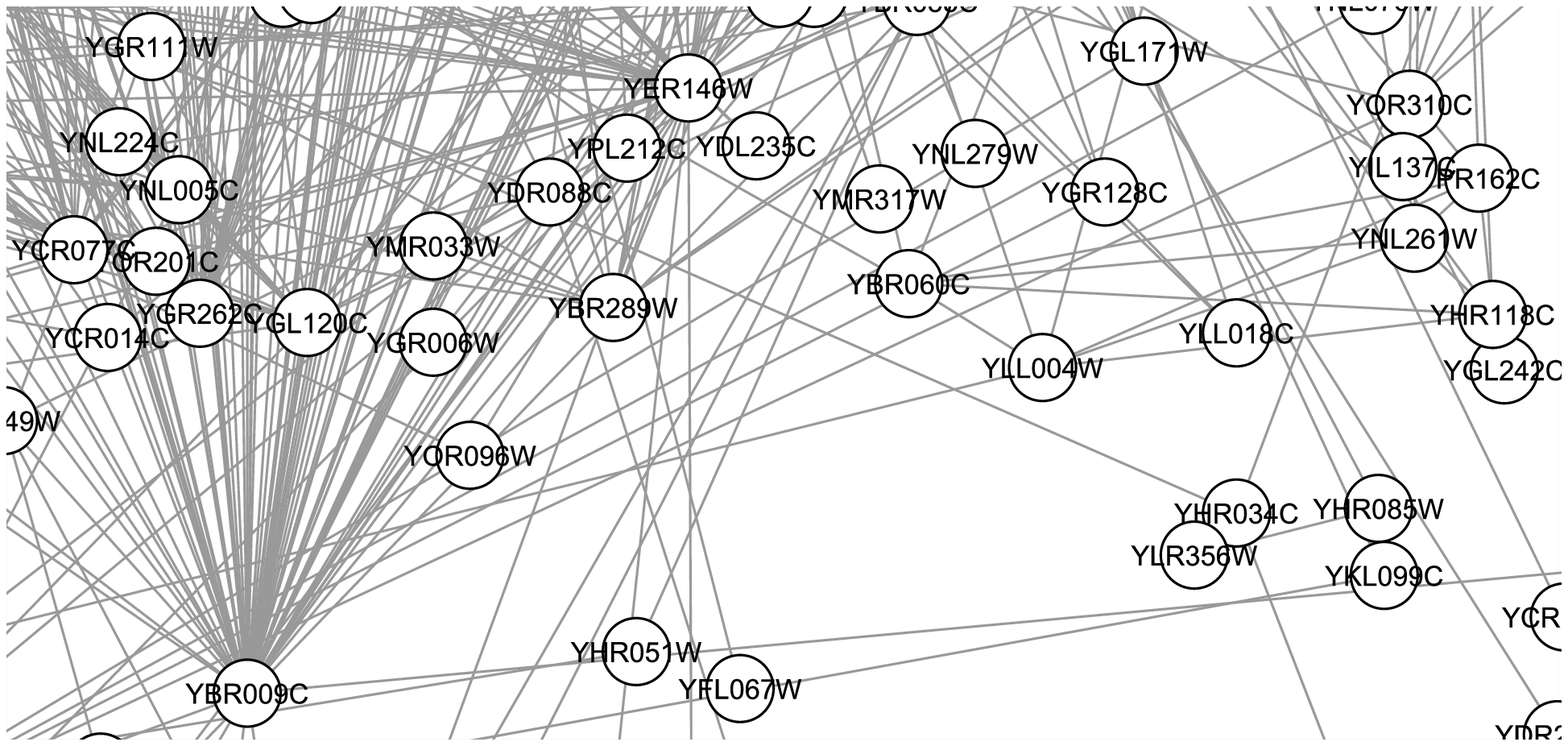}
\end{minipage}
  \caption{Global view of the protein interaction network of
{\it Saccharomyces cerevisiae} (left) comprising $14,319$ interactions over
$4389$ proteins, and a zoomed view of some proteins (right). A dot or circle
represents a protein and a line joins two interacting proteins. The protein
networks in this paper are drawn using the Cytoscape software \cite{Cyt03}.}
  \label{fig:yeastglobal}
\end{figure}

\paragraph{Making sense of it all.}
  Protein-protein interactions are the basis of many processes that maintain
cellular structure and function. Organizing the myriad protein interactions in
a cell into functional modules of proteins is a natural step in exploring
these cellular processes \cite{Har+99}. The need for such organization is
readily evident from a glance at the network of interactions observed in the 
yeast {\it Saccharomyces cerevisiae} (Figure \ref{fig:yeastglobal}). 
  Numerous such interactions are publicly available for other organisms too
from high-throughput though noisy experiments or extensive literature scanning
\cite{PIN04}. Computational methods are valuable to make sense of this raw
data and cope with its scale. These methods are similar in spirit to ones that
organize raw genomic sequences into functional elements like genes, regulatory
sequences, etc.

  In this work, a protein (interaction) network refers to a graph whose nodes
are the proteins of an organism and edges indicate physical interactions
between proteins, and a protein module refers to a subset of proteins in this
network along with the interactions between them. A functional module is then a
protein module known or supposed to be involved in a common cellular process
(eg.~a protein complex or a signaling pathway). Note however that concrete
definition of a cellular process is a subject of continuing discussion in
biology \cite{Har+99}. 

\paragraph{Previous work.} 
  Several methods analyse a single organism's protein network to identify
functional modules (see review \cite{PIN04}). A typical single-species method
uses connectivity information to cluster a protein network into highly
connected modules (eg.~MCODE \cite{BH03}). The method is effective if the
modules it outputs align well with a set of reference or known functional
modules, and if the output modules that don't match any known functional module
suggest plausible and testable biological hypotheses. 

  A few recent methods compare protein networks from two or more species to
identify functionally similar (conserved) protein modules between them (see
review \cite{SI06}). These pairwise or multiple network comparison methods
improve over single-species methods by using information on conservation
(cross-species similarity of protein sequences and interaction patterns) as
well as connectivity of the networks. 
  For instance amidst noisy data, conservation could reinforce evidence that
some connected proteins participate in a common function. These comparative
methods also enable transfer of functional annotations between organisms at the
level of conserved modules, interactions and proteins, a key utility
single-species methods cannot provide.  Similarly, conserved modules could
provide a basis for studies on the evolution of cellular structures and
networks. 
 
  Current network comparison methods include NetworkBLAST \cite{Sha+05}, MaWISh
\cite{KGS05}, a Bayesian method \cite{BL06} and Gr\ae mlin \cite{Fla+06}.
At a high level, these methods formulate a biologically-inspired measure to
score when a set of subgraphs from the input networks constitute a conserved
module and use a greedy heuristic to search for all high-scoring similar
subgraphs between the networks. Heuristics are necessary as their complicated
scoring measures lead to intractable (NP-hard \cite{GJBook}) search problems. 

\paragraph{This work.} 
  We present a pairwise network comparison method based on a graph-matching
algorithm with provable guarantees. Our novel graph-matching algorithm and the
guarantees on both its correctness and running time makes this work markedly
different from previous methods relying on search heuristics. Our search
formulation is biologically meaningful and yields promising results in
detecting functional modules and transferring functional annotations.  

  We formulate a conserved protein module as a pair of connected and locally
matching subsets of proteins, one from each input network. A locally matching
subset pair comprises protein pairs that have similar sequences and similar
neighborhood or context in the networks. Search for such conserved modules is
simpler than search formulations in previous methods and hence admits an
efficient polynomial-time algorithm. The main operation of this algorithm is a
recursive match and split of the proteins in the two input networks. We assess
the statistical significance of the conserved modules found by the algorithm
using similar paths statistics and reliabilities of noisy interactions.

  Our polynomial-time search problem is a novel contribution in the broader
field of graph-matching as well. Graph-matching refers to a class of problems
that find similar subgraphs between two graphs (see \cite{Sch05} for other
classes). Many graph-matching problems in literature are NP-hard \cite{GJBook},
permitting only heuristic or approximate solutions, due to a stringent global
structural match that they require between the similar subgraphs, as opposed to
the local match our problem requires. 
  For instance, the maximum common subgraph problem requires an exact
isomorphic match between subgraphs and is NP-hard \cite{GJBook}. Problems that
require inexact match between graphs to deal with error-prone data are NP-hard
too for a family of graph edit cost functions \cite{Bun99}. Some studies use a
scoring function to measure how similar two subgraphs are. They find
high-scoring subgraph pairs by finding heavy-weight subgraphs in a product
graph of the input graphs \cite{Sha+05,KGS05}, which is also NP-hard by
reduction from the maximum weight induced subgraph \cite{KGS05} or maximum
clique problem \cite{GJBook}.

\section{Methods}

\subsection{Conserved module premise}
  Our method compares protein networks from two species to find functionally
similar or conserved protein modules between them. A conserved module is
intuitively a pair of protein modules that share cross-species similarity at
the node level (homology of corresponding protein sequences) and graph
structure level (pattern of interactions). Our method's specific {\em premise}
is that a conserved module is a pair of {\em connected} and {\em locally
matching} subsets of proteins, one from each input graph. Two proteins locally
match if their sequences and neighborhood in the network are similar, and a
collection of such protein pairs is a locally matching subset pair. We support
this premise in this section and formalize it in the next. 

  Our premise is biologically motivated. The premise's connectivity criterion
makes it likely for a subset of proteins to be functionally homogeneous and its
local matching criterion makes it likely for two protein subsets to be
functionally similar. 
  Moreover, these criteria are minimal requirements and hence sensitive in
detecting reference functional modules. To illustrate, a functional module's
counterparts in two different species needn't match exactly in their graph
structure due to evolutionary divergence or errors in the interaction data,
which makes a local match more sensitive than a stringent structural match like
exact isomorphism. The two criteria yield good specificity too with some
additional improvements detailed in Section \ref{sec:overall-search}. Note that
a method's sensitivity denotes the fraction of reference modules it detects and
specificity the fraction of modules output by it that match some reference
module. 

  Our premise is also computationally attractive as it results in a search
problem that admits provably good algorithms for many choices of the
connectivity and local matching criteria. As mentioned before in Section
\ref{sec:intro}, our method's tractable search formulation contrasts it from
previous network comparison methods and graph-matching problems. 

\subsection{Graph-matching engine} 
  Finding conserved modules under the premise just outlined reduces to a
graph-matching problem of finding similar subgraphs between two input graphs.
We present this graph-matching problem variant and our polynomial-time
algorithm for it in this section, keeping formal details to a minimum for
reader convenience (see Supplemental text for complete analysis). We
also discuss the algorithm's generality, which makes it relevant for
application areas beyond protein networks. 

\begin{figure}[thb]
  \input{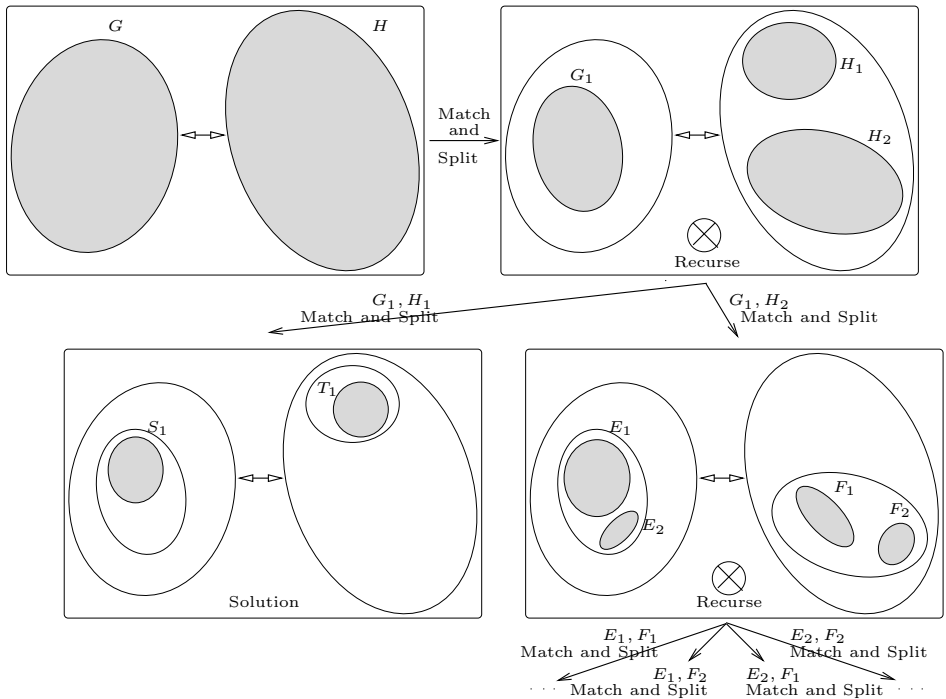}
  \caption{Pictorial sketch of the main operations of our graph-matching
algorithm. We show the input graphs $G,H$ and their subgraphs as ovals, hiding
node and edge details. 
  The algorithm focuses only on the shaded subgraph pairs at any point in its
execution, and refines them recursively until all solutions (similar subgraph
pairs) are found. A refinement involves doing a match and a split step to
compute locally matching and connected node-sets between two shaded subgraphs
(see text for details). 
  The subgraph pair $S_1,T_1$ is a solution, and the algorithm might find more
solutions as it recurses on the subgraph pairs $E_i,F_j$.  The statistically
significant solutions are finally output as conserved modules.}
  \label{fig:algodemo}
\end{figure}

  As a prelude, we sketch our graph-matching engine in Figure
\ref{fig:algodemo} and informally outline it now in the context of protein
networks. Say we start with the yeast and human protein networks along with the
homologous protein pairs between them. Our algorithm first computes {\em
locally matching} proteins between the two networks, and safely discards the
other proteins (i.e., any yeast protein with no human homolog or with some
human homolog but with poor local match, and vice versa). The algorithm next
splits the remaining yeast and human networks into {\em connected} sets of
proteins. 
  These connected subgraphs are locally matching with respect to the full input
networks and contain amongst them the actual solutions with refined local
matching relations. To find the solutions, our algorithm repeats the above
match and split steps on each pair of the connected subgraphs recursively (as
in Figure \ref{fig:algodemo}). The ensuing text provides precise descriptions
on matching general graphs.

\subsubsection{Problem statement} \label{sec:problem}
  We are given as input two graphs and a node similarity function {\em
sim}$(.,.)$. The function {\em sim}$(u,v)$ is true whenever node $u$ is similar
to node $v$ (eg.~based on sequence similarity of proteins) and false
otherwise.  This {\em sim}$(.,.)$ is a symmetric function defined over all
pairs of nodes $u,v$, one from each input graph. The problem now is to list
pairs of connected and locally matching subgraphs between the input graphs. In
this work, a subgraph of a graph usually refers to an induced subgraph, which
is a subset of nodes in the graph along with all edges between them. 

\begin{figure}[thb]
  \input{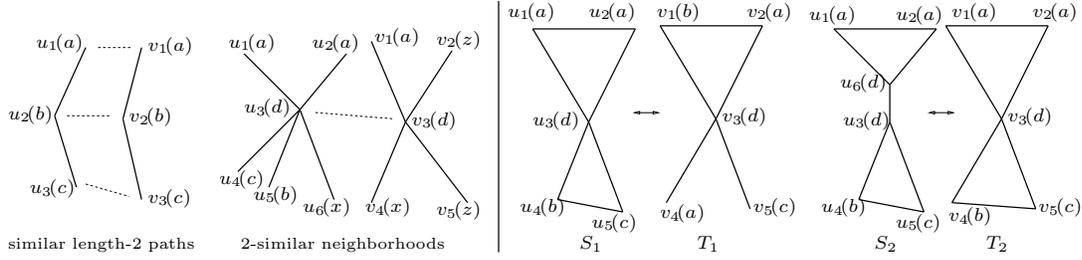}
  \caption{Illustration of two similar local structures
(left) and two solutions (right). Assume {\em sim}$(u_i,v_j)$ is true whenever
$u_i,v_j$ have the same label shown inside brackets. The dotted lines (left)
show some of the locally matching node pairs. The subgraph pair $S_1,T_1$ is a
solution when the local matching criterion is based on similar length-$p$ paths
($p\!=\!1 \mbox{ or } 2$) or $1$-similar neighborhoods, and $S_2,T_2$ is a
solution when the criterion is based also on $2$-similar neighborhoods.}
  \label{fig:localmatch}
\end{figure}

  We first build a {\em local-match}$_{S,T}(u,v)$ function for any subgraph
pair $S,T$ of the input graphs using the {\em sim}$(.,.)$ function. This new
function captures local or contextual match between the nodes $u$ of $S$ and
$v$ of $T$ using similar local structures present around these nodes in $S$ and
$T$. 
  Two possible choices of similar local structures are: {\em similar length-$p$
paths} for some small $p$ (say $2$), and {\em $s$-similar neighborhoods} around
nodes for some small $s$ (see Figure \ref{fig:localmatch}). In the former case,
{\em local-match}$_{S,T}(u,v)$ is true whenever some length-$p$ path in $S$
containing $u$ is similar to some length-$p$ path in $T$ containing $v$ (two
paths are similar if all their corresponding nodes are similar according to
{\em sim}$(.,.)$). In the latter case, {\em local-match}$_{S,T}(u,v)$ is true
whenever {\em sim}$(u,v)$ is true and {\em sim}$(u',v')$ is true for at least
$s$ distinct neighbor pairs $u'$ of $u$ in $S$ and $v'$ of $v$ in $T$. The
stringency of this criterion increases with $s$, with $1$-similar neighborhoods
being the least stringent. 

  We are ready to state the problem. Given two input graphs $G, H$, a node
similarity function {\em sim}$(.,.)$, and a {\em local-match}$_{S,T}(.,.)$
function computable for any two subgraphs $S,T$, the problem is to find all
maximal induced subgraph pairs $S \subseteq G$, $T \subseteq H$ that satisfy
two criteria: 

\smallskip \noindent
\hspace{0.1in}\begin{minipage}[h]{0.98\textwidth} 
{\em Connectivity:} $S, T$ are each connected, and \\
{\em Local Matching:} Each node $u$ in $S$ locally matches at least one node
$v$ in $T$ according to the {\em local-match}$_{S,T}(u,v)$ function, and
vice versa.  
\end{minipage}

\smallskip \noindent
Any subgraph pair that satisfy the above two criteria is called a solution (see
Figure \ref{fig:localmatch}), and maximality requires that of two solutions $S,
T$ and $S', T'$ with $S' \subseteq S, T' \subseteq T$, we only output the
maximal one $S, T$ to avoid redundancy. 
 
\subsubsection{Algorithm and analysis} \label{sec:algoanals}
  We present a simple and efficient algorithm for the above problem for any
{\em monotone} local matching criterion. A monotone criterion is one where any
two nodes that locally match remain so even after adding more nodes to the
subgraphs under consideration (i.e., if {\em local-match}$_{S,T}(u,v)$ is true,
then {\em local-match}$_{S',T'}(u,v)$ is also true for any $S' \supseteq S, T'
\supseteq T$). The similar length-$p$ paths or $s$-similar neighborhoods
criterion from last section are monotone. A useful property of monotonicity is
that the maximal solutions are only quadratic in number, since it lets us merge
any two solutions $S,T$ and $S',T'$ with a common node pair $u,v$ (i.e., $u \in
S \cap S', v \in T \cap T'$) into one solution $S \cup S', T \cup T'$. 
 
  Our algorithm presented below matches and splits the nodes of G and H into
smaller components, and then recurses on each of the component pairs. For
induced subgraphs $S \subseteq G, T \subseteq H$, we let $lm(S, T)$ denote all
nodes $u$ in $S$ for which {\em local-match}$_{S,T}(u,v)$ is true for some node
$v$ in $T$. 

\medskip \noindent
\begin{minipage}[h]{\textwidth}
{\em Match-and-Split}($G, H$): \\
  {\em [Match]} Compute induced subgraph:  

  \noindent \hspace{2em}
  $G'$ of $G$ over the locally matching nodes $lm(G, H)$, and 

  \noindent \hspace{2em}
  $H'$ of $H$ over the locally matching nodes $lm(H, G)$.\\
  {\em [Split]} Find connected components: 

  \noindent \hspace{2em} $G_1, \ldots, G_c$ of $G'$, and 

  \noindent \hspace{2em} $H_1, \ldots, H_d$ of $H'$.\\
  {\em [Recurse]} \\
  {\bf if} ($c = 1, d = 1$ and $G' = G, H' = H$)

  \noindent \hspace{1em} Output the maximal solution $G, H$. [base case]\\
  {\bf else} 

  \noindent \hspace{1em} {\bf for} $i = 1 \mbox{ to } c, j = 1 \mbox{ to } d$

  \noindent \hspace{2em} {\em Match-and-Split}($G_i, H_j$). [recursive case]\\ 
\end{minipage}

\vspace{-1em}
\paragraph{Correctness.} 
  Consider any solution $S,T$. Each recursive call retains all locally matching
nodes and processes all pairs of resulting components. Hence in at least one
path of the recursion call tree, all nodes in $S,T$ remain locally matching due
to monotonicity and connected as part of a bigger subgraph pair. This retained
unsplit $S,T$ is finally output as part of a solution. Further, the output
solutions are maximal because no node pair is common to any two output
solutions (as shown in the Supplemental text). 

\paragraph{Running time.} 
  Let $n_F, m_F$ denote the number of nodes, edges respectively in a graph $F$.
The algorithm runs in time $O(n_G n_H + (n_G + n_H) m_G m_H)$ on the graphs
$G,H$ when the local matching criterion is similar \mbox{length-$1$} paths.
This running time bound follows from bounds proved in the Supplemental text,
which hold for more general monotone local matching criteria. The algorithm is
efficient in practice too as the locally matching proteins between two graphs
in our experiments reduces drastically as the recursion depth increases. 

\subsubsection{Generality of the algorithm} \label{sec:generality} 
  Our algorithm framework admits quite general schemes to search for similar
subgraph pairs and score them, and is hence attractive in a biological
setting. The searching scheme is flexible as our problem variant and algorithm
works for different connectivity and local matching criteria. The scoring
scheme is flexible as it is decoupled from the searching scheme. Besides, the
number of maximal solutions is only quadratic, so it is not expensive to
compute a sophisticated, biologically-inspired score for every solution. 

  We discuss the flexibility of the local matching and connectivity criteria in
more detail. We already saw different monotone local matching criteria. We
could also combine such monotone criteria to get a new monotone criterion. For
example, declaring two nodes as locally matching if they are so with respect to
similar length-$p$ paths {\em or} $s$-similar neighborhoods gives a less
stringent criterion. Many connectivity criteria are possible too. We could
replace connectedness with biconnectedness \cite{BaaBook} for instance by
simply changing the algorithm's split step, and still obtain a provably
efficient algorithm (see Supplemental text). 

\subsection{Overall method} \label{sec:overall} 
  Our method of detecting conserved modules between two protein networks
involves a searching scheme to find similar subgraph pairs ({\em candidate
conserved modules} or {\em candidates}) using our graph-matching algorithm
above, and a scoring scheme to rank these candidates using statistics of
similar paths between a subgraph pair. We now describe these schemes and their
place in the overall {\em Match-and-Split} method. 

\subsubsection{Searching scheme (via graph-matching)} \label{sec:overall-search}
  To adapt the generic graph-matching algorithm {\em Match-and-Split} to the
specific task of comparing protein networks, we make default choices for
certain graph-matching parameters and incorporate a clustering heuristic to
handle solutions that are large. In this section, our algorithm's maximal
solutions are referred to simply as solutions. 

\paragraph{Choosing parameters.}
  Our default parameter choices result in a lenient graph-matching criterion,
for we would like to detect as many functional modules as possible from noisy
protein networks of divergent organisms (eg.~yeast and human). The default
choices we made on exploring a limited parameter space follow. 

  Connectedness defines our connectivity criterion. We choose it over
biconnectedness as some functional modules (eg.~linear signaling pathways) are
not biconnected, and even those over highly interacting proteins (eg.~protein
complexes) may not appear biconnected due to incomplete interaction data. 

  Similar length-$p$ paths (for $p\!=\!1,2$) defines our local matching
criterion. We choose it over $s$-similar neighborhoods (for $s\!=\!1,2$) again
on the basis of sensitivity. For $p,s\!=\!1$, both options are equivalent as
they yield the same $lm(S, T)$ node-set (defined in Section
\ref{sec:algoanals}).  However for $p,s\!=\!2$, this node-set from similar
paths is a superset of the one from similar neighborhoods, so similar paths is
a more lenient criterion. 

  Sequence similarity defines our node similarity function as in previous
network comparison methods. We in fact choose the same criteria used in two
previous methods for fair evaluation. The two criteria are: ($A$) {\em
sim}$(u,v)$ is true whenever the BLAST E-value of proteins $u, v$ is at most
$10^{-7}$ and each protein is among the $10$ best BLAST matches of the other
\cite{Sha+05}, and ($B$) {\em sim}$(u,v)$ is true whenever the BLAST E-value of
$u, v$ is less than that of $60\%$ of ortholog pairs in some ortholog database
(see \cite{Koy+06} for details). 

\paragraph{Incorporating clustering heuristic.} 
  Increased sensitivity from the lenient graph-matching criterion above comes
at a cost. Sometimes, a solution is over a large number of proteins and hence
less specific. For instance, a solution from a preliminary comparison of yeast
and human networks covers more than $500$ yeast and human proteins! 
  To split such large solutions, we incorporate a betweenness clustering
heuristic in our {\em Match-and-Split} algorithm. This clustering splits a
graph into highly-connected, smaller clusters based on iterative computations
of an edge betweenness centrality measure \cite{GN02} (see Supplemental text
for details). One could also cluster a graph using other methods such as the
popular spectral clustering methods \cite{Wei99}. 

  We incorporate the clustering by replacing our algorithm's `[base-case]'
statement with the code block below. As before $n_G$ refers to the number of
nodes in $G$, and we may assume $n_G \ge n_H$ without loss of generality. The
parameter $n_{\max}$ (say $25$) indicates when a solution is large. 

\medskip \noindent
  [base case] code block:\\
  {\bf if} ($n_G \le n_{\max}$ and $n_H \le n_{\max}$) 

  \noindent \hspace{1em} Output the maximal solution $G, H$.\\
  {\bf else} [large solution]

  \noindent \hspace{1em} Split $G$ into clusters $G_1,\ldots,G_e$ using betweenness clustering. 

  \noindent \hspace{1em} {\bf for} $i = 1 \mbox{ to } e$

  \noindent \hspace{2em} {\em Match-and-Split}($G_i, H$). 

\medskip \noindent
  A betweenness clustering of $G$ takes $O(n_G m_G^2)$ running time
\cite{GN02}. In practice, incorporating the clustering heuristic is not
expensive as our experiments result in very few large solutions (mostly one),
each covering just a few hundred nodes. 

\subsubsection{Scoring scheme (via similar paths)} \label{sec:overall-score}
  We score a candidate conserved module based on the number of similar
length-$p$ paths, and express the statistical significance of the score as a
P-value. We use the P-values both to rank the candidates from the searching
scheme and to retain only those with P-values at $10\%$ significance level
after multiple testing. 
  Our scoring scheme is flexible, as seen in Section \ref{sec:generality}, in
permitting complicated scoring measures including measures in previous network
comparison methods. Still we use a simple scoring measure and the promising
results we obtain shows the strength of our searching scheme based on the
graph-matching algorithm. 

  The score of a candidate conserved module $S \subseteq G, T \subseteq H$,
where $G,H$ are the input protein networks, is simply the number of pairs of
similar length-$p$ paths between them (defined in Section \ref{sec:problem}).
We evaluate the P-value of this score using a null model that randomizes the
edges and node similarity function of $G, H$ to exclude the mechanism of
interest viz., conservation of protein modules. 
  To provide a stringent control, the randomization loosely preserves the
degree sequence and node similarity distribution as in previous methods. The
simplicity of our scoring measure and null model allows us to develop an
analytical bound on the P-value (see Supplemental text). This bound can also
incorporate reliabilities of noisy protein interactions. 

\subsubsection{Implementation pipeline and dataset} \label{sec:overall-impl}
  The overall Match-and-Split method proceeds in a pipeline to detect candidate
conserved modules between two protein networks. First our searching scheme uses
the novel graph-matching algorithm to produce candidates, then a size filter
retains only medium-sized candidates, and finally our scoring scheme ranks the
candidates by their P-values and retains those at $10\%$ significance level
(after multiple testing). A fast implementation of the method is publicly
available (see Supplemental text for website reference), and it takes only a
minute or few (at most $4$) on a $3.4$ GHz Pentium Linux machine to compare two
studied networks.

  The size filter, similar to one in a previous method \cite{Sha+05}, retains
any candidate subgraph pair $S,T$ whose number of nodes $n_S, n_T$ satisfy
$n_{\min} \le n_S,n_T \le n_{\max}$ ($n_{\min} = 3, n_{\max} = 25$ in our
experiments, and $n_{\max}$ is same as in the searching scheme's clustering
heuristic). We focus on such medium-sized candidates for the following reasons.
A large module as a whole is likely to correspond to a less specific function
and worse causes artifactual increase in sensitivity in our evaluation studies.
A small module, over say two proteins, is likely to result from a spurious
match occurring simply by chance. 

  The protein networks for model organisms {\it Saccharomyces cerevisiae}, {\it
Drosophila melanogaster} and {\it Caenorhabditis elegans} (referred hereafter
as yeast, fly and worm respectively) are experimentally-derived (eg.
two-hybrid, immunoprecipitation) interactions collected in the DIP database
\cite{DIP04}. The version of the networks used and the interaction
reliabilities based on a logistic regression model are taken from a previous
study \cite{Sha+05}. Our human protein network is from the HPRD database
\cite{HPRD03} (binary or direct interactions; Sep 2005 version), and we assume
unit reliability for these interactions as they are literature-based. 

\subsection{Evaluation measures} \label{sec:evaln} 
  We describe some measures that we would need in the Results section to
evaluate and interpret the conserved modules from pairwise network
comparisons.  The notation yeast-human comparison denotes the comparison
between yeast and human protein networks, and yeast-human modules the resulting
conserved modules. Similar notations apply for other two-species comparisons
too. 

\subsubsection{Performance (sensitivity and specificity)} \label{sec:evaln-perf}
  The performance of a pairwise network comparison method depends on the
quality of candidate conserved modules it outputs, which we could measure by
how well the candidate set aligns with a reference set of conserved modules.
But such reference sets, even if available, are not comprehensive or applicable
for our purpose (eg.~STKE (\url{http://stke.sciencemag.org/cm/}, Jun 2005)
contains only a couple of fly-worm conserved signaling pathways and no protein
complexes; Biocarta (\url{http://www.biocarta.com/genes/allPathways.asp}, Jan
2006) has many human-mouse conserved pathways but available mouse interaction
data is sparse). 

  A viable alternative is to use known functional modules in a {\em single
species} as reference to evaluate part of a candidate set. Our reference
modules are the literature-based yeast protein complexes present in MIPS
\cite{MIPS04} (May 2006 version), at level at most $3$ in its complex category
hierarchy. We consider only complexes with $3$ to $25$ proteins due to our
study's focus on medium-sized modules (see Section \ref{sec:overall-impl} for
reasons). 
  To test a method, we compare the yeast network against the human, fly or worm
network (dataset details in Section \ref{sec:overall-impl}), collect the yeast
subgraphs $S$ from each output candidate $S,T$, and measure the overlap of
these single-species candidate modules with the reference modules. 

  Our main measures are module-level sensitivity and specificity, which are the
fraction of reference modules covered by some candidate module and that of
candidate modules covered by some reference module respectively. A module $S$
of proteins from a single species (yeast) is covered by another module $S'$ if
$|S \cap S'| / |S| \ge 50\%$, a stringent criterion given the noisy interaction
data. 

  We also present measures at the interaction and protein levels to assess the
quality of different components of candidate modules (similar to measures at
the gene, exon and nucleotide levels in gene structure prediction methods
\cite{BG96}). Define the protein interactions {\em spanned} by a set of modules
as the union of interactions present in each of these modules. If set $A$
denotes the interactions spanned by all candidate modules and $B$ that by all
reference modules, then interaction-level sensitivity is $|A \cap B| / |B|$ and
specificity is $|A \cap B| / |A|$. Protein-level measures are defined
similarly. 

\subsubsection{GO analysis measures} \label{sec:evaln-go}
  We use the GO resource \cite{GO00} for functional annotations of genes.
Specifically, all our functional descriptions are based on known GO Biological
Process annotations (Aug 2006 version) of proteins to terms at level at least
$4$ in the GO hierarchy. Given these annotations, we next define many
GO-related concepts that we would need later. 

  The {\em best GO term} of a protein module is the term the module's proteins
are enriched for with the least hypergeometric P-value (computed by
GO::TermFinder \cite{GOTermFinder} at $10\%$ significance level with Bonferroni
correction).  Given two GO terms, the {\em match} between them is the overlap
$\min(|A \cap B| / |A|, |A \cap B| / |B|)$ between the set $A,B$ of terms they
imply, where a GO term implies itself and its ancestors in the GO ontology.
This match is based on a well-justified measure \cite{KMF05}. 

  We call a protein module {\em functionally homogeneous} if at least $50\%$ of
its proteins are annotated with the best GO term the whole module is enriched
for. We call a candidate conserved module $S,T$ (eg.~yeast-human candidate) as
{\em functionally similar} with respect to GO if $S$ and $T$ are each
functionally homogeneous and the match between their best GO terms is at least
$75\%$. 

  When validating a predicted GO term of a protein, we use a well-justified
criterion as for match between terms. This criterion requires the respective
sets $A, B$ of terms implied by the predicted term and some known term
annotated to the protein to have a high overlap $|A \cap B| / |A| \ge 75\%$.

\section{Results}

\subsection{Performance against previous methods}
  We test our Match-and-Split method against two previous methods, NetworkBLAST
and MaWISh. We try two Match-and-Split versions, $p\!=\!1$ and $p\!=\!2$, which
specify the local matching criterion of similar length-$p$ paths. NetworkBLAST
has two versions too, a clusters one to detect conserved protein complexes and
a paths one to detect conserved paths of length just $4$. We don't test the
Bayesian \cite{BL06} and Gr\ae mlin \cite{Fla+06} methods as they are from very
recent studies with results focusing on coexpression or prokaryotic networks,
whereas the current study's focus is eukaryotic protein networks.
Fundamentally though, all these methods work for the networks of any species. 

  We evaluate a method by measuring how well the yeast modules it outputs (on
pairwise protein network comparisons of yeast and other species) align with a
reference set of yeast protein complexes in MIPS, as explained in Section
\ref{sec:evaln-perf}. 
  For fair evaluation, we attempt to use common input, default parameter
values, and common output processing for all methods. For instance, the input
networks {\em and} node similarity function {\em sim}$(.,.)$ are the same
across methods. The $3,25$ size filter thresholds are same too, so we evaluate
all methods only on the {\em medium-sized candidates} they output (see Section
\ref{sec:overall-impl}). 

  Different {\em sim}$(.,.)$ criteria could yield quite varying results, and we
try two criteria used in previous methods as discussed in Section
\ref{sec:overall-search}. The main text presents criterion $A$ results as every
method has better or comparable module-level results with criterion $A$ than
$B$. The Supplemental text has some criterion $B$ results to show few
departures from this trend (mainly in yeast-worm comparison and minor increase
of MaWISh specificity in yeast-human case) and indicate slight changes in the
relative performance of the methods between the two criteria. 

  First, we discuss module-level results of Match-and-Split ($p\!=\!1$)
relative to the other two methods. In yeast-human comparison (Table
\ref{tab:perf-sh}), Match-and-Split and MaWISh have comparable performance with
Match-and-Split being somewhat better. A similar trend appears in the yeast-fly
case (Table \ref{tab:perf-sd}) and with an exception in the yeast-worm case
(Table \tabperfsc). 
  In all three cases, NetworkBLAST (clusters) has better sensitivity than other
methods at the cost of very low specificity. NetworkBLAST (paths) achieves
better sensitivity and reasonable specificity, but only by outputting too many
and too small ($4$-protein) candidates. The small size permits a brute-force
search for optimal candidates, which is better than the greedy heuristic search
for larger candidates in NetworkBLAST (clusters). In contrast, the search in
Match-and-Split is based on a provably efficient graph-matching algorithm with
no candidate size restrictions. 

  Moving from relative to absolute performance, the values of module-level
sensitivity and specificity are low (for example, a mere $25.0\%$ and $47.5\%$
respectively by Match-and-Split ($p\!=\!1$), a competitive method in
yeast-human case). Low sensitivity could be due to factors like noisy
interaction data, poor conservation of complexes across compared organisms, or
differing computational and biological definitions of a functional module or
complex.  Finding which of these factors is key is a subject of future work.
Low specificity is probably due to incompleteness of the reference set, and it
does not indicate the output candidates are spurious (a claim supported by
further analysis in Section \ref{sec:transfer}). 

\subsection{Single-species vs pairwise network analysis}
  Pairwise network comparison methods use cross-species conservation in an
attempt to improve over single-species methods in detection of functional
modules (as seen in Section \ref{sec:intro}). But previous network comparison
studies have not evaluated their methods against single-species ones. Here we
undertake this evaluation by testing a popular single-species method MCODE
\cite{BH03} on the yeast network under the same measures and size range ($3$ to
$25$ proteins) as before. It is beyond the scope of this study to test all
single-species methods. 

  The performance of the pairwise methods relative to the single-species MCODE
is varied. Under proper homolog and species selection, Match-and-Split performs
better than MCODE in detecting reference yeast complexes. Comparing Table
\ref{tab:perf-sh} on yeast-human comparison with Table \ref{tab:perf-mcode},
Match-and-Split ($p\!=\!1$) is much more sensitive than MCODE at the same
specificity, and MaWISh performs similar to MCODE. 
  The choice of homologs is important as changing the {\em sim}$(.,.)$
criterion from $A$ here to $B$ in Table \tabperfcritB~reduces the benefit of
pairwise methods over MCODE. Species selection is also crucial as the pairwise
methods have worse sensitivity than MCODE in the yeast-fly (Table
\ref{tab:perf-sd}) and yeast-worm (Table \tabperfsc) cases. Low sensitivity in
the yeast-worm case could be due to the sparse worm interaction data, but the
reason in the yeast-fly case is unclear as the fly network is interaction-rich. 

\begin{table}[t]
\begin{tabular}{|l| p{2.699cm} ||c|c||c|c|c|c|}\hline
 Method & {\small \#}$\!$ output modules & \multicolumn{2}{c||}{Module} & \multicolumn{2}{c|}{Interaction} & \multicolumn{2}{c|}{Protein} \\\cline{3-8}
  & {\small (interactions, proteins)} & Sens. & Spec. & Sens. & Spec. & Sens. & Spec. \\\hline
Match-and-Split & & & & & & & \\
{\small ($p\!=\!1$)} & 80 {\small (667, 421)} & 25.0 & 47.5 & 20 & 35 & 25 & 48 \\
{\small ($p\!=\!2$)} & 72 {\small (664, 411)} & 28.0 & 41.7 & 20 & 34 & 25 & 47 \\\hline
NetworkBLAST & & & & & & & \\
{\small (clusters)} & \ 66 {\small (1042, 522)} & 32.6 & 18.2 & 30 & 33 & 33 & 50 \\
{\small (paths, length 4)} & 276 {\small (625, 563)} & 28.0 & 48.2 & 18 & 33 & 34 & 48 \\\hline
MaWISh & 151 {\small (508, 389)} & 20.5 & 43.7 & 15 & 33 & 23 & 46 \\\hline
\end{tabular}
\caption{Evaluation of output candidates from yeast-human network comparison
using sensitivity (sens.) and specificity (spec.) measures (expressed as
rounded percentages) at the module, interaction and protein levels. The second
column shows the number of yeast modules (candidates) output, and the number of
interactions and proteins spanned by these yeast modules. The reference set
comprises $132$ medium-sized (size 3 to 25) yeast complexes in MIPS that span
$1144$ interactions and $791$ proteins. All relevant definitions are in Section
\protect{\ref{sec:evaln-perf}}.}
\label{tab:perf-sh} 
\end{table}

\begin{table}[t]
\begin{tabular}{|l| p{2.7cm} ||c|c||c|c|c|c|}\hline
 Method & {\small \#}$\!$ output modules & \multicolumn{2}{c||}{Module} & \multicolumn{2}{c|}{Interaction} & \multicolumn{2}{c|}{Protein} \\\cline{3-8}
  & {\small (interactions, proteins)} & Sens. & Spec. & Sens. & Spec. & Sens. & Spec. \\\hline
Match-and-Split & & & & & & & \\
{\small ($p\!=\!1$)} & 27 {\small (155, 123)} & ~6.8 & 51.9 & ~5 & 35 & ~8 & 49 \\
{\small ($p\!=\!2$)} & 25 {\small (131, 110)} & ~7.6 & 48.0 & ~4 & 37 & ~7 & 53 \\\hline
NetworkBLAST & & & & & & & \\
{\small (clusters)} & 71 {\small (626, 442)} & 12.1 & ~2.8 & 10 & 19 & 19 & 34 \\
{\small (paths, length 4)} & 165 {\small (296, 405)} & ~7.6 & 27.3 & ~5 & 18 & 17 & 34 \\\hline
MaWISh & 26 {\small (81, 87)} & ~5.3 & 50.0 & ~3 & 40 & ~6 & 52 \\\hline
\end{tabular}
\caption{Evaluation of output candidates from yeast-fly network comparison,
using the same format as Table \protect{\ref{tab:perf-sh}}.}
\label{tab:perf-sd}
\end{table}

\begin{table}[t!]
\begin{tabular}{|l| p{2.7cm} ||c|c||c|c|c|c|}\hline
 Method & {\small \#}$\!$ output modules & \multicolumn{2}{c||}{Module} & \multicolumn{2}{c|}{Interaction} & \multicolumn{2}{c|}{Protein} \\\cline{3-8}
  & {\small (interactions, proteins)} & Sens. & Spec. & Sens. & Spec. & Sens. & Spec. \\\hline
MCODE & 53 {\small (614, 323)} & 16.7 & 47.2 & 19 & 36 & 20 & 48 \\\hline
\end{tabular}
\caption{Evaluation of output clusters from yeast network analysis by MCODE, a
single-species clustering method, using the same format as Table
\protect{\ref{tab:perf-sh}}. We focus on medium-sized (size 3 to 25) clusters
as in other evaluations.} 
\label{tab:perf-mcode}
\end{table}

  Our Match-and-Split searching scheme includes a betweenness clustering
heuristic. The heuristic is by itself a single-species method (denoted
Split-only) for it can cluster the full yeast network into highly connected
modules. Table \ref{tab:perf-bcc} compares Match-and-Split and Split-only to
show the boost in specificity from adding pairwise match criterion to plain
clustering. Split-only detects more reference yeast complexes but at the cost
of outputting too many candidate modules at very low specificity. Adding
pairwise match criterion also drastically reduces running time by restricting
the size of graphs that need to be betweenness clustered. 

\clearpage

\begin{table}[thb]
\begin{tabular}{|l| p{2.7cm} ||c|c||c|c|c|c|}\hline
 Method & {\small \#}$\!$ output modules & \multicolumn{2}{c||}{Module} & \multicolumn{2}{c|}{Interaction} & \multicolumn{2}{c|}{Protein} \\\cline{3-8}
  & {\small (interactions, proteins)} & Sens. & Spec. & Sens. & Spec. & Sens. & Spec. \\\hline
Match-and-Split & 81 {\small (634, 410)} & 24.2 & 48.2 & 20 & 35 & 25 & 49 \\\hline
Split-only & 569 {\small (3597, 2776)} & 50.0 & 12.3 & 58 & 18 & 76 & 22 \\\hline
\end{tabular}
\caption{Evaluation of Match-and-Split ($p\!=\!1$) on pairwise yeast-human
network comparison against Split-only on single-species yeast network
clustering, again using the same format as Table \protect{\ref{tab:perf-sh}}.
Similar results from Match-and-Split ($p\!=\!2$) is omitted. 
  As betweenness clustering of large graphs is compute-intensive, Split-only
uses a quicker version of it (see Supplemental text; Split-only still requires
orders of magnitude more time than Match-and-Split). For fairness, the
Match-and-Split version here uses the quicker clustering inside its searching
scheme.}  
\label{tab:perf-bcc}
\end{table}

\begin{figure}[thb]
\begin{minipage}{.45\textwidth}
  \begin{center}
    \includegraphics[width=.9in,height=!]{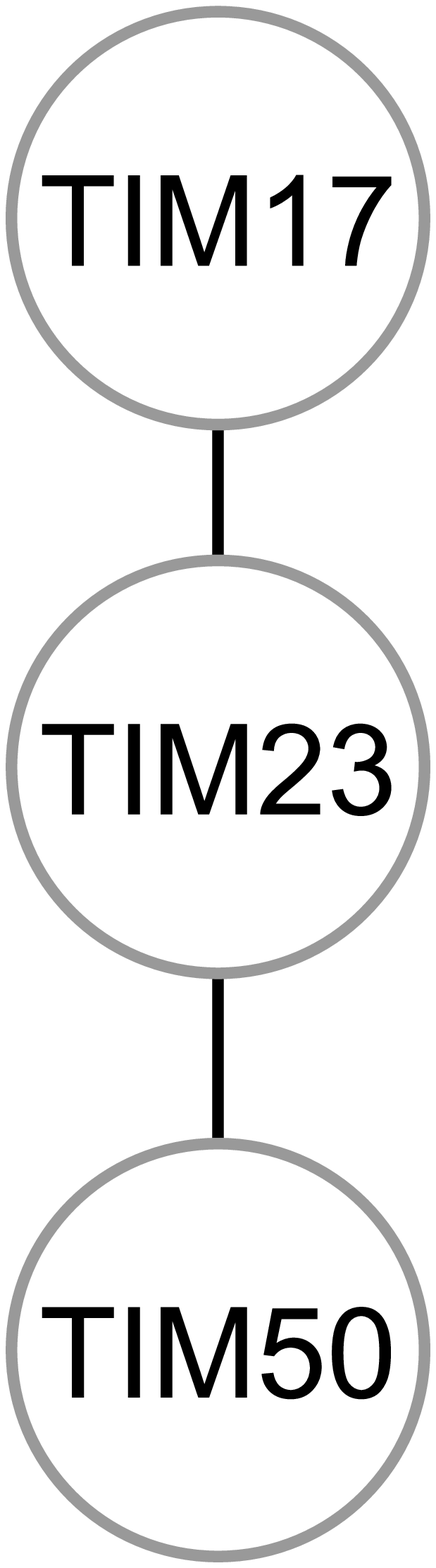}
  \end{center}
\end{minipage}
\hfill
\begin{minipage}{.45\textwidth}
  \begin{center} 
    \includegraphics[width=.9in,height=!]{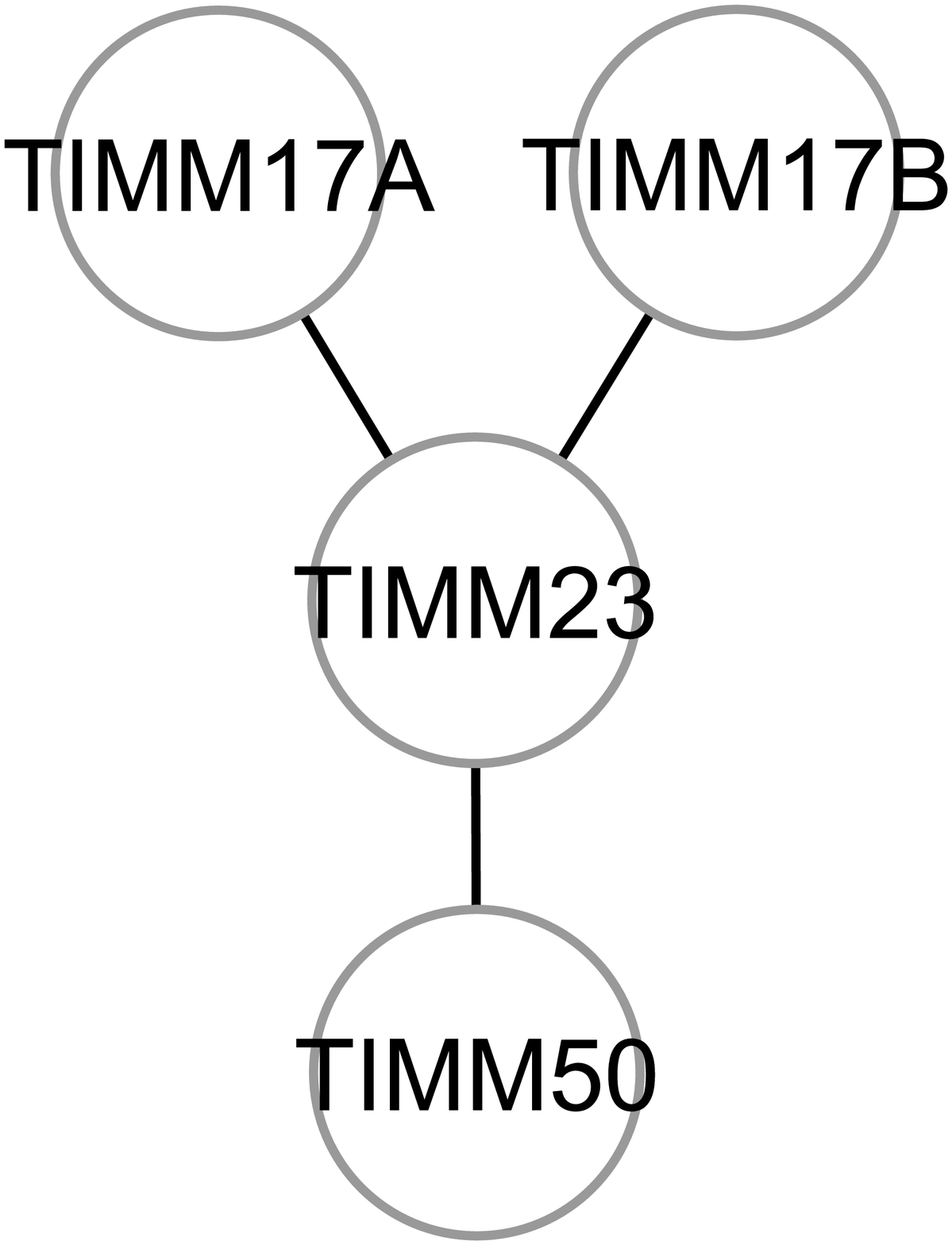}
  \end{center}
\end{minipage}
\hrule
\begin{minipage}{.45\textwidth}
  \begin{center} 
    \includegraphics[width=1.5in,height=!]{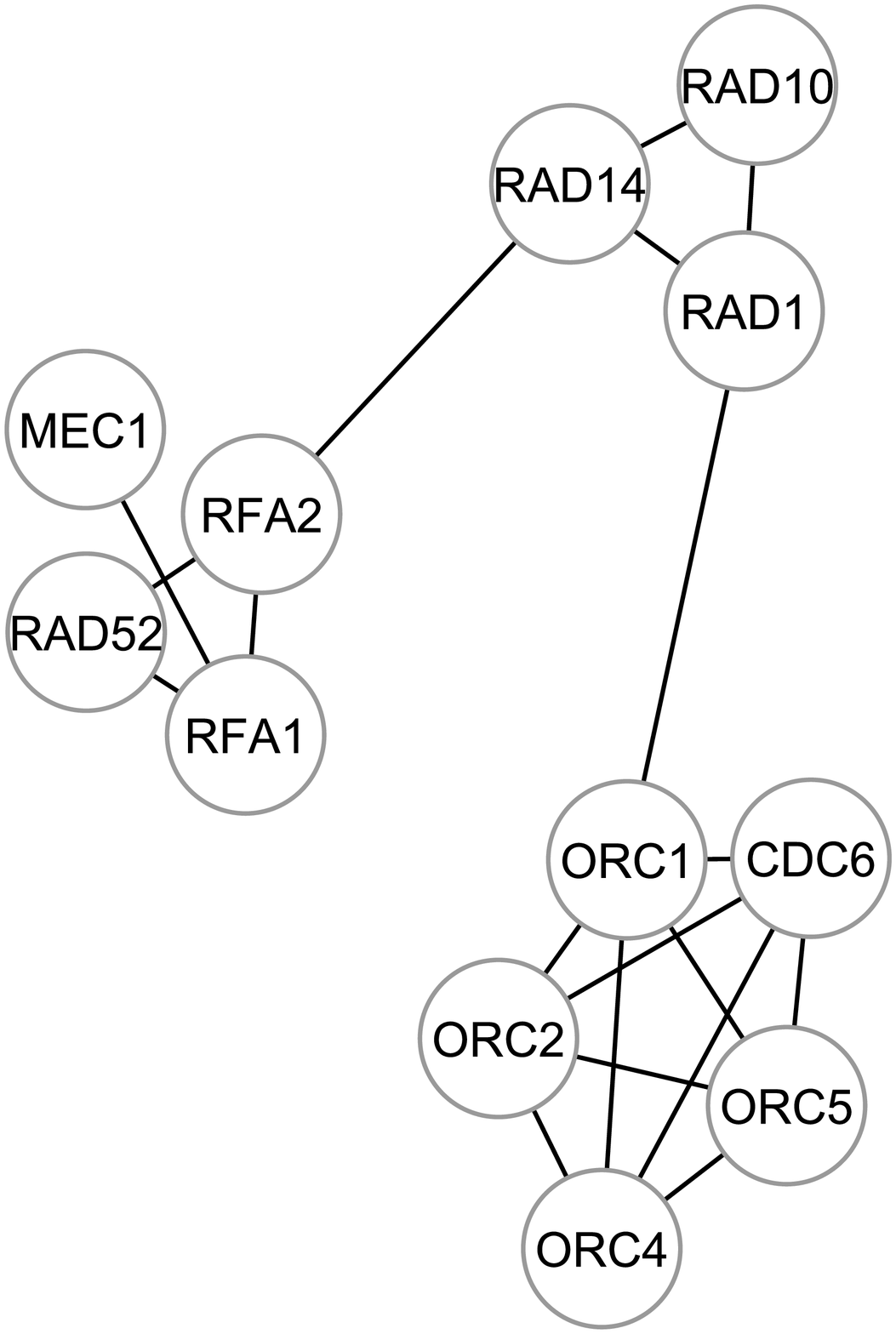}
  \end{center}
\end{minipage}
\hfill
\begin{minipage}{.45\textwidth}
  \begin{center} 
    \includegraphics[width=1.5in,height=!]{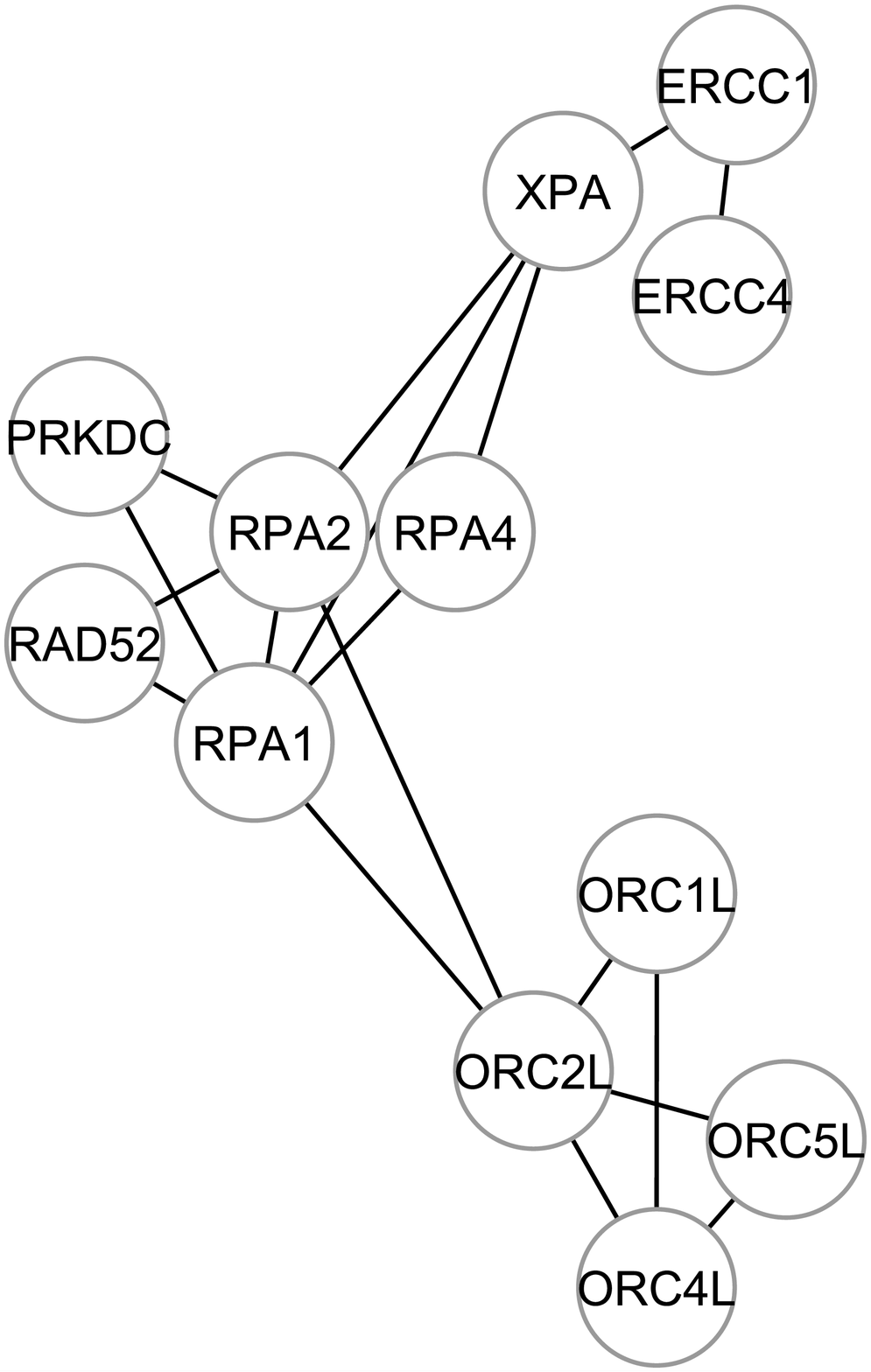}
  \end{center}
\end{minipage}
\caption{Select candidates from Match-and-Split ($p\!=\!1$)
yeast-human comparison. Each candidate is a conserved module of yeast (left)
and human (right) proteins. Two proteins similar by the {\em sim}$(.,.)$
function are roughly aligned horizontally.}
\label{fig:select-sh}
\end{figure}

\subsection{Select conserved modules}
  The results above evaluate the candidates from our method on a global scale.
Here we discuss the biology of a select few candidates. We start with a flavour
of some top-ranked candidates from the Match-and-Split ($p\!=\!1$) yeast-human
comparison in Tables \ref{tab:top5-sh} and \ref{tab:top5sz-sh}. These tables
contain functional descriptions based on some known GO Biological Process
annotations as described in Section \ref{sec:evaln-go}. The format of these
tables is inspired from a previous study \cite{KGS05}. 

  Consider the candidate ranked $2$ in Table \ref{tab:top5-sh} and shown in
Figure \ref{fig:select-sh}. From literature-based descriptions in SGD
\cite{SGD} and UniProt \cite{UniProt04}, the yeast and human proteins of this
candidate are each components of the TIM23 complex, a mitochondrial inner
membrane translocase. This complex mediates translocation of preproteins across
the mitochondrial inner membrane. Typical preproteins are nuclear-encoded,
synthesized in the cytosol and contain a targeting sequence (presequence or
transit peptide) to direct transport. 

  We now elaborate on the candidate ranked $72$ in Table \ref{tab:top5sz-sh}
and shown in Figure \ref{fig:select-sh}. The candidate may seem too
heterogeneous to be conserved but it actually contains many homologous
complexes as inferred from literature-based comments at SGD and UniProt. This
example illustrates how a lenient matching criterion over noisy interactions
could detect known complexes. 
  The origin recognition complex (of the ORC proteins) with counterparts in
yeast and human binds replication origins, and plays a role in DNA replication
and transcriptional silencing. The RAD1, RAD10 and RAD14 proteins are subunits
of the Nucleotide Excision Repair Factor 1 (NEF1) in yeast and homologous to
the ERCC4, ERCC1 and XPA respectively in human. The RFA1, RFA2 in yeast
(homologs RPA1, RPA2 in human) are subunits of heterotrimeric Replication
Factor A (RF-A), a single-stranded DNA-binding protein involved in DNA
replication, repair and recombination. 

\begin{table}[tb]
\begin{tabular}{|l|l|c| p{4cm} | p{4cm} | p{1.3cm} |}\hline
 Ra- & P-value & Size & \multicolumn{3}{c|}{Best GO term of module (\% annotated proteins)} \\\cline{4-6}
 nk & (score) & & Yeast & Human & Terms' match \\\hline
1 & 3.33e-13 (8) & 5, 3 & purine ribonucleoside salvage (100\%) & nucleoside metabolism (100\%) & 53\% \\\hline
2 & 1.05e-12 (3) & 3, 4 & protein import into mitochondrial matrix (100\%) & protein targeting to mitochondrion (75\%) & 89\% \\\hline
3 & 1.38e-12 (4) & 3, 3 & postreplication repair (100\%) & DNA repair (100\%) & 94\% \\\hline
4 & 1.40e-12 (6) & 4, 3 & ER to Golgi vesicle-mediated transport (100\%) & ER to Golgi vesicle-mediated transport (100\%) & 100\% \\\hline
5 & 1.89e-12 (2) & 3, 3 & processing of 20S pre-rRNA (100\%) & rRNA processing (100\%) & 95\% \\\hline
\end{tabular}
\caption{Five top-ranked candidates from Match-and-Split ($p\!=\!1$)
yeast-human comparison. The size of a candidate (third column) is the number of
its yeast, human proteins.  The `\% annotated proteins' is the fraction of
proteins in a module annotated with the module's best GO term, and the match
shown is between the best GO terms of yeast module $S$ and human module $T$ in
a candidate $S,T$ (see Section \protect{\ref{sec:evaln-go}} for definitions).}
\label{tab:top5-sh}
\end{table}

\begin{table}[tb]
\begin{tabular}{|l|l|c| p{4cm} | p{4cm} | p{1.3cm} |}\hline
 Ra- & P-value & Size & \multicolumn{3}{c|}{Best GO term of module (\% annotated proteins)} \\\cline{4-6}
 nk & (score) & & Yeast & Human & Terms' match \\\hline
50 & 7.76e-10 (40) & 12, 10 & ubiquitin-dependent protein catabolism (100\%) & ubiquitin-dependent protein catabolism (100\%) & 100\% \\\hline
72 & 1.02e-08 (16) & 12, 12 & DNA-dependent DNA replication (75\%) & DNA metabolism (91.7\%) & 83\% \\\hline
74 & 1.48e-08 (95) & 16, 17 & protein amino acid phosphorylation (81.2\%) & phosphorus metabolism (88.2\%) & 35\% \\\hline
77 & 6.40e-08 (18) & 15, 13 & transcription initiation (86.7\%) & transcription initiation (69.2\%) & 100\% \\\hline
78 & 1.28e-07 (79) & 20, 23 & actin filament organization (65\%) & Rho protein signal transduction (43.5\%) & 11\% \\\hline
\end{tabular}
\caption{Five top-ranked candidates with at least $10$ yeast and $10$ human
proteins from Match-and-Split ($p\!=\!1$) yeast-human comparison, presented in
the same format as Table \protect{\ref{tab:top5-sh}}.}
\label{tab:top5sz-sh}
\end{table}

\subsection{Annotation transfer from yeast to human} \label{sec:transfer}
  The candidates output by pairwise network comparison methods, like the ones
sampled in last section, enable transfer of functional annotations between
organisms. The idea is to annotate a protein module in one organism with a
function that a similar module in another organism is known to be enriched
for.  Our focus here is annotation transfer from yeast to human based on
candidate conserved modules (i.e., yeast-human candidates) output by
Match-and-Split ($p\!=\!1$), using certain known GO Biological Process
annotations described in Section \ref{sec:evaln-go}. The Match-and-Split
results in this section are competitive with the corresponding results of other
tested methods (shown in Tables \tabgofnsshANDtabtransfersh). 

  We first present results on functional homogeneity and similarity of
yeast-human candidates (as defined with respect to GO in Section
\ref{sec:evaln-go}). Collect the yeast module $S$ of every yeast-human
candidate $S,T$. Then all ($100\%$) of these yeast modules are homogeneous for
some function, which could then be transferred. This fraction is $83.8\%$ for
the human modules collected similarly. The fraction of yeast-human candidates
functionally similar with respect to GO is a reasonable $42.5\%$, which further
supports annotation transfer. 

  The actual transfer on each yeast-human candidate $S,T$ involves assigning
all human proteins in $T$ to the best GO term the yeast module $S$ is enriched
for. If this procedure predicts more than one GO term for a human protein, we
retain the term with the least hypergeometric P-value (described in Section
\ref{sec:evaln-go}). We predict GO Biological Process terms for a total of
$462$ human proteins (predictions available online; see Supplemental text). 
  This annotation transfer is reliable as a reasonable $295$ of these
predictions are valid by a stringent, well-justified criterion in Section
\ref{sec:evaln-go}. This transfer covers only $462$ proteins though, a small
fraction of the $1882$ proteins in the human network sequence-similar to some
protein in the yeast network (by the {\em sim}$(.,.)$ function).

\section{Discussion}
  In the context of comparing two protein networks, this work shows it is
possible to design a provably good search algorithm that also translates to
promising performance in practice. The algorithmic guarantees of our
Match-and-Split method distinguishes it from previous methods based on greedy
heuristics. Formal guarantees are important as they lend credibility to the
conserved modules found by a bounded-time search procedure. 

  Our method Match-and-Split performs competitively in tests against two
previous pairwise network comparison methods. For instance, Match-and-Split
performs comparable to or somewhat better than the MaWISh method. Relative to
NetworkBLAST, Match-and-Split detects less reference complexes but at much
higher specificity. In tests against a single-species method MCODE,
Match-and-Split performs better in yeast-human comparison. This single-species
test, which was not done in previous pairwise studies, also reveals comparisons
(yeast-fly and yeast-worm) where the pairwise methods are poorer than MCODE. 

  The above evaluations, especially the single-species one, leads to an
immediate future question. Are the poor results of pairwise methods in some
comparisons mainly due to incomplete interaction data or something intrinsic to
the choice of the species pair and homologs between them? The answer could
inform the conditions when pairwise methods exploit cross-species conservation
to improve single-species detection of functional modules. 

  The conserved modules our method detects and their functional descriptions
would be the findings of most practical interest to biologists. Also of
interest would be the reasonably accurate functional predictions resulting from
the transfer of GO annotations between conserved modules. We make these
findings publicly available (at a website mentioned in the Supplemental text),
along with a fast Match-and-Split implementation to facilitate new network
comparisons. 

  Our graph-matching algorithm is flexible in allowing diverse local matching
and connectivity criteria. A biologist could for instance design a stringent
matching criterion to detect similar instances of a functional module in a
single-species network (duplicated modules). We could even compare other types
of networks like metabolic networks within the same algorithmic framework. A
challenge then is the judicious design of a matching criterion for the
biological comparison of interest. 

  Our algorithm guarantees are limited to monotone local matching criteria.
Many useful criteria, such as one that declares two nodes locally matching if
they are similar and at least half of their neighbors are similar, are
non-monotone. A future direction is to explore tractable search formulations
for non-monotone criteria. 
  Another limitation with the current study, but not with our framework, is the
use of a simple scoring scheme. The simple scores yield reasonably good
results, however network conservation scores that correlate better with the
biological significance of a conserved module is a subject of future work.

\section*{Acknowledgments}
  We thank Jayanth Kumar Kannan for valuable discussions that led to the
algorithm. We thank Sourav Chatterji, Sridhar Rajagopalan and Ben Reichardt for
comments at different stages of the project. We thank Roded Sharan, Silpa
Suthram and Mehmet Koyut\"urk for help with their previous work. This work was
partially supported by NSF Award 331494.

\bibliographystyle{plain}
\bibliography{paper}

\end{document}